\documentclass[submission]{eptcs}
\providecommand{\event}{GanDALF 2011}

\usepackage{amssymb,amsmath,amsthm,amscd,latexsym,graphicx}
\usepackage{gastex}
\usepackage{times}

\newcommand{\slopefrac}[2]{\leavevmode\kern.1em
  \raise .5ex\hbox{\the\scriptfont0 #1}\kern-.1em
  /\kern-.15em\lower .25ex\hbox{\the\scriptfont0 #2}}
\newcommand{\half}{\slopefrac{1}{2}}
\newcommand{\dist}{{\mathcal{D}}} 
\newcommand{\trans}{\delta} 
\newcommand{\VE}{V_{E}}
\newcommand{\VA}{V_{A}}
\newcommand{\VR}{V_{R}}
\newcommand{\set}[1]{\{ #1 \}}
\newcommand{\seq}[1]{\langle #1 \rangle}
\newcommand{\size}{\mathrm{size}}
\newcommand{\co}{\mathrm{co}}
\newcommand{\UP}{\mathrm{UP}}
\newcommand{\NP}{\mathrm{NP}}
\newcommand{\play}{\pi} 
\newcommand{\Play}{\Pi} 
\newcommand{\strateve}{\sigma} 
\newcommand{\stratadam}{\tau} 
\newcommand{\finitewords}[1]{#1^*} 
\newcommand{\Strateve}{\Sigma} 
\newcommand{\Stratadam}{\Gamma} 
\newcommand{\A}{\mathcal{A}} 
\newcommand{\Pa}{\mathcal{P}}
\newcommand{\Reach}{\mathrm{Reach}}
\newcommand{\Inf}{\mathrm{Inf}}
\newcommand{\Parity}{{\mathrm{Parity}}}
\newcommand{\MP}{\mathrm{MeanPayoff}}
\newcommand{\DP}{\mathrm{DiscPayoff}}
\newcommand{\N}{\mathbb{N}}
\newcommand{\R}{\mathbb{R}}
\newcommand{\Q}{\mathbb{Q}}
\newcommand{\C}{\mathcal{C}}
\newcommand{\Pro}{\mathbb{P}}
\newcommand{\val}[1]{\langle{#1}\rangle}
\newcommand{\valeve}{\val{\mathrm{E}}}
\newcommand{\valadam}{\val{\mathrm{A}}}
\newcommand{\Exp}{\mathbb{E}}
\newcommand{\vw}{\mathrm{v_{win}}}
\newcommand{\vl}{\mathrm{v_{lose}}}
\newcommand{\Red}{\mathfrak{R}}
\newcommand{\hpi}{\widehat{\play}}
\newcommand{\Reachgen}[2]{\Reach(#1,#2)}
\newcommand{\Crossgen}[1]{\mathrm{Cross}(#1)}
\newcommand{\Rpath}[1]{\mathrm{ReachPath}(#1)}
\newcommand{\Rloop}[2]{\mathrm{ReachLoop}(#1,#2)}
\newcommand{\Cpath}{\mathrm{CrossPath}}
\newcommand{\Cloop}[1]{\mathrm{CrossLoop}(#1)}

\newtheorem{theorem}{Theorem}
\newtheorem{property}{Property}

\def\qed{\rule{0.4em}{1.4ex}}

\title{A reduction from parity games to simple stochastic games
\thanks{The research was supported by Austrian Science Fund (FWF) NFN Grant S11407-N23 (RiSE) and a Microsoft faculty fellowship.}}
\author{
Krishnendu Chatterjee
\institute{IST Austria (Institute of Science and Technology, Austria)}
\email{krishnendu.chatterjee@ist.ac.at}
\and
Nathana\"el Fijalkow 
\institute{IST Austria (Institute of Science and Technology, Austria)}
\institute{\'ENS Cachan (\'Ecole Normale Sup\'erieure de Cachan, France)}
\email{nathanael.fijalkow@gmail.com}
}
\def\titlerunning{From parity games to simple stochastic games}
\def\authorrunning{K. Chatterjee \& N. Fijalkow}

\begin{document}

\maketitle

\begin{abstract}
Games on graphs provide a natural model for reactive non-terminating systems.
In such games, the interaction of two players on an arena
results in an infinite path that describes a run of the system.
Different settings are used to model various open systems in computer science,
as for instance turn-based or concurrent moves,
and deterministic or stochastic transitions.
In this paper, we are interested in turn-based games,
and specifically in deterministic parity games and stochastic reachability games
(also known as simple stochastic games).
We present a simple, direct and efficient reduction from deterministic parity games
to simple stochastic games:
it yields an arena whose size is linear up to a logarithmic factor 
in size of the original arena.
\end{abstract}

\noindent{\bf Keywords.} 
\emph{Stochastic games, parity objectives, reachability objectives.}

\section{Introduction}
\medskip\noindent{\bf Graph games.}
Graph games are used to model reactive systems.
A finite directed graph, whose vertices represent states 
and edges represent transitions, models the system.
Its evolution consists in interactions between a controller and the environment,
which is naturally turned into a game on the graph between two players,
Eve and Adam.
In the turn-based setting, in each state of the system,
either the controller chooses the evolution of the system
(the corresponding vertex is then controlled by Eve),
or the system evolves in an uncertain way, 
then aiming at the worst-case scenario Adam controls the corresponding vertex.
This defines a $2$-player arena as a finite directed graph and
a partition of the vertex set into Eve and Adam vertices.
However, in many applications, systems are randomized,
leading to the definition of stochastic arenas:
in addition to Eve and Adam vertices,
the graph also has random vertices
where the evolution is chosen according to a given probability distribution.

A pebble is initially placed on the vertex representing 
the initial state of the system, 
then Eve, Adam and random move this pebble along the edges,
constructing an infinite sequence of vertices.
The sequence built describes a run of the system: 
Eve tries to ensure that it satisfies some specification of the system,
while Adam tries to spoil it.

\medskip\noindent{\bf Parity objectives.}
The theory of graph games with $\omega$-regular winning conditions is the 
foundation for modelling and synthesizing reactive processes with fairness
constraints.
The {\em parity\/} objectives provide an adequate model, as the fairness constraints 
of reactive processes are $\omega$-regular, and every $\omega$-regular 
winning condition can be specified as a parity objective~\cite{Tho97}.
We consider $2$-player games with parity objectives:
deciding the winner in polynomial time is a longstanding open question,
despite many efforts from a large community.
The best known upper-bound is $\UP \cap \co\UP$~\cite{Jur98}.

\medskip\noindent{\bf Simple stochastic games.}
Considering probabilistic games instead of deterministic 
allows the description much more
reactive systems by modelling uncertainty, 
but leads to higher complexity for corresponding decision problems.
We consider stochastic games with reachability objectives:
a given vertex is distinguished, and Eve tries to reach it.
Those games were introduced by Condon, and named 
simple stochastic games~\cite{Con92-IC}.
We consider the following decision problem: 
can Eve ensure to reach the target vertex
with probability more than half?
As for the above decision problem, the best known upper-bound is 
$\NP \cap \co\NP$~\cite{Con92-IC}.

\medskip\noindent{\bf Reduction: from parity games to simple stochastic games.} 
The notion of reduction between games is an important aspect in the study 
of games as it allows to understand which classes of games are 
subsumed by others.
A classical reduction of $2$-player parity games to simple stochastic 
games is through a sequence of three reductions: 
(a)~from $2$-player parity games to $2$-player mean-payoff (or limit-average)
games~\cite{Jur98}; 
(b)~from $2$-player mean-payoff games to $2$-player discounted-payoff 
games~\cite{ZP96}; and 
(c)~from $2$-player discounted-payoff games to stochastic 
reachability games~\cite{ZP96}.
The sequence of reductions yields the following result: given a $2$-player 
parity game with $n$ vertices, $m$ edges, and a parity objective with $d$ 
priorities, the simple stochastic game obtained through the 
sequence of reductions has $n+m$ vertices,
including $m$ probabilistic ones, $4\cdot m$ edges and the size of the arena is 
$O(m \cdot d \cdot \log(n))$.

\medskip\noindent{\bf Our results:} 
we present a direct reduction of $2$-player parity games to 
simple stochastic games, and thus show that one can discount the step of 
going through mean-payoff and discounted games.
Moreover, our reduction is more efficient: given a $2$-player 
parity game with $n$ vertices, $m$ edges, and a parity objective with $d$ 
priorities, the simple stochastic game obtained by our direct reduction
has $n+m$ vertices among which $m$ are probabilistic, 
$3 \cdot m$ edges and the size of the arena is $O(m\cdot \log(n))$.
Finally, we conclude following proof ideas from~\cite{Con93-DMTCS} 
that the decision problem for simple stochastic games is in $\UP \cap \co\UP$, 
and from~\cite{AM09,CH08} we obtain that the decision problems in 
stochastic parity, mean-payoff and discounted games 
all are in $\UP \cap \co\UP$.

\section{Definitions}

Given a finite set $A$, a probability distribution $\mu$ 
on $A$ is a function $\mu: A \to [0,1]$ such that $\sum_{a \in A} \mu(a)=1$.
We denote by $\dist(A)$ the set of all probability distributions on $A$.

\medskip\noindent{\bf Stochastic arena.} 
A stochastic (or $2\half$-player) arena $G = ((V,E),(\VE,\VA,\VR),\trans)$
consists of a finite directed graph $(V,E)$ with vertex set $V$ and edge set $E$,
a partition $(\VE,\VA,\VR)$ of the vertex set $V$
and a probabilistic transition function $\trans: \VR \to \dist(V)$ that given a 
vertex in $\VR$ gives the probability of transition to the next vertex.
Eve chooses the successor of vertices in $\VE$,
while Adam chooses the successor of vertices in $\VA$;
vertices in $\VR$ are random vertices and their successor is chosen 
according to $\trans$. 
We assume that for all $u \in \VR$ and $v \in V$ we have 
$(u,v) \in E$ if and only if $\trans(u)(v) > 0$.
We assume that the underlying graph has no deadlock: every vertex has a successor.
The special case where $\VR = \emptyset$ corresponds to $2$-player arenas 
(for those we omit $\trans$ from the description of the arena).

\smallskip\noindent{\bf Size of an arena.} 
The size of a stochastic arena $G = ((V,E),(\VE,\VA,\VR),\trans)$ is 
the number of bits required to store it:
$$\size(G) = \underbrace{\log(n)}_{\textrm{vertices}} + 
\underbrace{2 \cdot m \cdot \log(n)}_{\textrm{edges}} + 
\underbrace{n + n_R \cdot \log(n)}_{\textrm{vertex partition}} \quad + 
\underbrace{\size(\trans)}_{\textrm{probabilistic transitions}}$$ 
where $n = |V|$, $m = |E|$, $n_R = |\VR|$ 
and $\size(\trans) = \sum_{u \in \VR} \sum_{v \in V} |\trans(u)(v)|$,
where $|\trans(u)(v)|$ is the length of the binary representation of 
$\trans(u)(v)$.

\medskip\noindent{\bf Plays and strategies.} 
A \emph{play} $\play$ in a stochastic arena $G$ is an 
infinite sequence $\seq{v_0,v_1,v_2,\dots}$ of vertices such that 
for all $i \geq 0$ we have $(v_i,v_{i+1}) \in E$.
We denote by $\Play$ the set of all plays. 
A \emph{strategy} for a player is a recipe that prescribes how to play, 
\textit{i.e}, given a finite history of play, a strategy defines the next move.
Formally, a strategy for Eve is a function 
$\strateve : \finitewords{V}\cdot \VE \to V$ 
such that for all $w \in \finitewords{V}$ and $v \in \VE$ 
we have $(v,\strateve(w \cdot v)) \in E$.
We define strategies for Adam analogously, 
and denote by $\Strateve$ and $\Stratadam$ the set of all strategies 
for Eve and Adam, respectively.
A strategy is \emph{memoryless}, or \emph{positional} 
if it is independent of the history of play and only depends on the current vertex,
\textit{i.e}, for all $w,w' \in \finitewords{V}$ 
and $v \in \VE$ we have $\strateve(w \cdot v) = \strateve(w' \cdot v)$.
Hence a memoryless strategy can be described as a function $\strateve: \VE \to V$.

Once a starting vertex $v \in V$ and strategies $\strateve$ for Eve 
and $\stratadam$ for Adam are fixed, the outcome 
of the game is a random walk $\play(v,\strateve,\stratadam)$ 
for which the probabilities of events are uniquely defined, where an \textit{event} 
$\A \subseteq \Play$ is a measurable set of plays. 
For an event $\A \subseteq \Play$, 
we write $\Pro^{\strateve, \stratadam}_v(\A)$ for the probability 
that a play belongs to $\A$ if the game starts from the vertex 
$v$ and the players follow the strategies $\strateve$ and $\stratadam$.
In case of $2$-player arenas, 
if we fix positional strategies $\strateve$, $\stratadam$,
and a starting vertex $v$, then the play $\play(v,\strateve,\stratadam)$ 
obtained is unique and consists in a simple path $\seq{v_0, v_1, \ldots v_{l-1}}$ 
and a cycle $\seq{v_l, v_{l+1}, \ldots, v_k}$ 
executed infinitely often, \textit{i.e}, the play is a ``lasso-play'':
$\seq{v_0, v_1, \ldots, v_{l-1}} \cdot \seq{v_l, v_{l+1}, \ldots, v_k}^\omega$. 

\medskip\noindent{\bf Qualitative objectives.}
We specify \emph{qualitative} objectives for the players by providing
a set of \emph{winning} plays $\Phi \subseteq \Play$ for each player.
We say that a play $\play$ {\em satisfies} the objective
$\Phi$ if $\play \in \Phi$.
We study only zero-sum games, where
the objectives of the two players are complementary,
\textit{i.e}, if Eve has the objective~$\Phi$, 
then Adam has the objective $\Play \setminus \Phi$.
\begin{itemize}
	\item \emph{Reachability objectives.}
Given a set $T\subseteq V$ of ``target'' vertices, the reachability
objective requires that some vertex of $T$ be visited.
The set of winning plays is 
$\Reach(T) = \set{\seq{v_0, v_1, v_2,\dots} \in \Play 
\mid v_k \in T \mbox{ for some }k \geq 0}$.
	\item \emph{Parity objectives.}
Let $p : V \to \N$ be a function 
that assigns a \emph{priority} $p(v)$ to every vertex $v \in V$.
For a play $\play = \seq{v_0, v_1, \dots} \in \Play$,
we define 
$\Inf(\play) = \set{v \in V \mid \mbox{$v_k = v$ for infinitely many $k$}}$
to be the set of vertices that occur infinitely often in~$\play$.
The \emph{parity objective} is defined as
$\Parity(p)= \set{\play \in \Play \mid \min(p(\Inf(\play))) \mbox{ is even}}$.
In other words, the parity objective requires that the minimum 
priority visited infinitely often is even.
\end{itemize}

\noindent{\bf Quantitative objectives.} A \emph{quantitative} objective
is specified as a measurable function $f : \Play \to \R$.
In zero-sum games the objectives of the players are functions $f$ and 
$-f$, respectively. 
We consider two classes of quantitative objectives, namely, 
mean-payoff and discounted-payoff objectives.
\begin{itemize}
	\item \emph{Mean-payoff objectives.} 
Let $r : V \to \R$ be a real-valued reward function that assigns to 
every vertex $v$ the reward $r(v)$.
The \emph{mean-payoff} objective $\MP(r)$ assigns to every play the ``long-run'' 
average of the rewards appearing in the play.
Formally, for a play $\play = \seq{v_0,v_1,v_2,\dots}$ we have
$$\MP(r)(\play) = \liminf_{n \to \infty} \frac{1}{n+1} \cdot \sum_{i=0}^n r(v_i)$$
	\item \emph{Discounted-payoff objectives.} 
Let $r : V \to \R$ be a reward function and $0<\lambda<1$ be a 
discount factor, the discounted-payoff objective $\DP(\lambda,r)$ 
assigns to every play the discounted sum of the rewards in the play. 
Formally, for a play $\play = \seq{v_0,v_1,v_2,\dots}$ we have
$$\DP(\lambda,r)(\play) = (1 - \lambda) \cdot \lim_{n \to \infty} 
\sum_{i = 0}^n \lambda^i \cdot r(v_i)$$
\end{itemize}

\medskip\noindent{\bf Values and optimal strategies.}
Given objectives $\Phi \subseteq \Play$ for Eve and 
$\Play \setminus \Phi$ for Adam, and 
measurable functions $f$ and $-f$ for Eve and Adam, 
respectively, we define the \emph{value} functions
$\valeve$ and $\valadam$ for Eve and Adam, respectively, as the following 
functions from the vertex space $V$ to the set $\R$ of reals:
for all vertices $v \in V$, let
$$\begin{array}{rclrcl}
\valeve(\Phi)(v) & = & 
\displaystyle \sup_{\strateve \in \Strateve} \inf_{\stratadam \in \Stratadam} 
\Pro_v^{\strateve,\stratadam}(\Phi); 
 &  \quad 
\valeve(f)(v) & = & 
\displaystyle \sup_{\strateve \in \Strateve} \inf_{\stratadam \in \Stratadam} 
\Exp_v^{\strateve,\stratadam}[f]; \\[2ex]
\valadam (\Play \setminus \Phi)(v) & = & 
\displaystyle \sup_{\stratadam\in \Stratadam} 
\inf_{\strateve \in \Strateve} \Pro_v^{\strateve,\stratadam}(\Play \setminus \Phi);
& \quad 
\valadam (-f)(v) & = &  
\displaystyle \sup_{\stratadam\in \Stratadam} 
\inf_{\strateve \in \Strateve} \Exp_v^{\strateve,\stratadam}[-f].
\end{array}$$

In other words, the values $\valeve(\Phi)(v)$ and $\valeve(f)(v)$ 
give the maximal probability 
and expectation with which Eve can achieve 
her objectives $\Phi$ and $f$ from vertex~$v$,
and analogously for Adam.
The strategies that achieve those values are called optimal:
a strategy $\strateve$ for Eve  is \emph{optimal} from the vertex
$v$ for the objective $\Phi$ if 
$\valeve(\Phi)(v) = \inf_{\stratadam \in \Stratadam} 
\Pro_v^{\strateve, \stratadam}(\Phi)$;
and $\strateve$ is \emph{optimal} from the vertex $v$ for $f$ if 
$\valeve(f)(v) = \inf_{\stratadam \in \Stratadam} \Exp_v^{\strateve,\stratadam}[f]$.
The optimal strategies for Adam are defined analogously.

\begin{theorem}[Memoryless determinacy~\cite{ZP96}]\label{memdet} 
For all stochastic arenas, 
\begin{enumerate}
	\item For all objectives $\Phi$ such that $\Phi$ is either a 
reachability or a parity objective, for all vertices $v$ we have
$$\valeve(\Phi)(v) + \valadam(\Play \setminus \Phi)(v) = 1.$$ 
Memoryless optimal strategies exist for both players from all 
vertices.
Furthermore, for the case of $2$-player arena, then for all $v \in V$, 
$\valeve(\Phi)(v) \in \set{0,1}$. 
	\item For all objectives $f:\Play \to \R$ such that $f$ is either 
a mean-payoff or discounted-payoff objective, for all vertices $v$ we have
$$\valeve(f)(v) + \valadam(-f)(v) = 0.$$ 
Memoryless optimal strategies exist for both players from all vertices.
\end{enumerate}
\end{theorem}

\medskip\noindent{\bf Games.}
A stochastic game is given by an arena and an objective.
As a special case, a $2$-player game is given by a $2$-player arena 
and an objective.
For instance, a $2$-player parity game is a couple $(G,\Parity(p))$,
where $G$ is a $2$-player arena,
and a stochastic reachability game is a couple $(G,\Reach(T))$,
where $G$ is a stochastic arena.

We define simple stochastic games to be special case of stochastic reachability game 
where $V$ contains two distinguished absorbing vertices $\vw$ and $\vl$ 
and the reachability set is $T = \set{\vw}$.
Once a play reached one of the two vertices $\vw$ or $\vl$,
the game is stopped as its outcome is fixed.
A simple stochastic game has the stopping property if 
for all strategies $\strateve$ and $\stratadam$ and all vertices $v \in V$, 
$\Pro^{\strateve,\stratadam}_v (\Reach\set{\vw,\vl}) = 1$.

\medskip\noindent{\bf Decision problems for games.}
Given an arena $G$, an objective $\Phi$, a starting vertex $v \in V$ 
and a rational threshold $q \in \Q$, 
the decision problem we consider is whether $\valeve(\Phi)(v) \geq q$.
It follows from Theorem~\ref{memdet} that in $2$-player arenas, 
with a parity objective $\Parity(p)$, 
for a vertex $v$ we have $\valeve(\Parity(p))(v) \in\set{0,1}$.
If the value is $1$, then we say that Eve is winning, otherwise Adam is winning.

\section{A direct reduction}

In this section, we present a direct reduction 
to show that determining the winner in $2$-player parity 
games can be reduced to the decision problem of simple stochastic games 
with the threshold $\frac{1}{2}$.
Specifically, from a $2$-player parity game $(G,\Parity(p))$ 
and a starting vertex $v$
we show how to construct in polynomial time
a stochastic arena $\Red(G)$ with a reachability objective $\Reach(\vw)$ such that
Eve is winning in $G$ from $v$ for the parity condition if and only if
$\valeve(\Reach(\vw))(v) \geq \frac{1}{2}$ in $\Red(G)$.

\medskip\noindent{\bf Construction of the stochastic arena.}
We now present the construction of a stochastic arena $\Red(G)$:
$$\Red(G) = ((V \uplus E \uplus \set{\vw,\vl}, E'),
(\VE \uplus \set{\vw}, \VA \uplus \set{\vl}, E), \trans) \quad,$$ 
where $\uplus$ denotes the disjoint union.
The set of edges and the transition function is as follows:
$$\begin{array}{rcl}
E'  & = & \set{(u, (u,v)) ; ((u,v), v) ; ((u,v), \vw) 
\mid (u,v) \in E, p(v) \textrm{ even}} \\
& \cup & \set{(u, (u,v)) ; ((u,v), v) ; ((u,v), \vl) 
\mid (u,v) \in E, p(v) \textrm{ odd}} \\
& \cup & \set{(\vw,\vw) ; (\vl,\vl)}
\end{array} \quad.$$

The transition function $\trans$ is defined as follows: 
if $p(v)$ is even, $\trans((u,v))(\vw) = P_v$ 
and $\trans((u,v))(v) = 1 - P_v$, if $p(v)$ is odd, $\trans((u,v))(\vl) = P_v$ and 
$\trans((u,v))(v) = 1 - P_v$. 
We will describe $P_v$ as reals in the interval $(0,1)$ 
satisfying certain conditions, and we will prove correctness of the reduction 
as long as the conditions are satisfied.

\begin{figure}
\begin{center}
\begin{picture}(150,52)(0,4)
	\gasset{Nw=8,Nh=8}
  
	\rpnode[arcradius=1](1)(10,35)(6,5){$u$}
	\rpnode[arcradius=1](2)(35,35)(6,5){$v$}
	\drawedge(1,2){}

	\put(50,49){\Large If $v$ is even}
	\rpnode[arcradius=1](3)(80,50)(6,5){$u$}
	\rpnode[polyangle=90](4)(100,50)(3,5){}
	\rpnode[arcradius=1](5)(120,50)(6,5){$v$}

	\node[Nh=1,Nw=1](lefteyewin)(113.5,36){}
	\node[Nh=1,Nw=1](righteyewin)(116.5,36){}
	\drawcurve[AHnb=0](113,34)(114,32.5)(115.5,32.5)(117,34)

	\put(104,52){$1 - P_v$}
	\put(104,39){$P_v$}
  	\node(6)(115,35){}
  	\drawedge(3,4){}
  	\drawedge(4,5){}
  	\drawedge(4,6){}

	\put(50,19){\Large If $v$ is odd}
	\rpnode[arcradius=1](7)(80,20)(6,5){$u$}
  	\rpnode[polyangle=90](8)(100,20)(3,5){}
	\rpnode[arcradius=1](9)(120,20)(6,5){$v$}
	\put(104,22){$1 - P_v$}
	\put(104,9){$P_v$}
  	\node(10)(115,5){}

	\node[Nh=1,Nw=1](lefteyewin)(113.5,6){}
	\node[Nh=1,Nw=1](righteyewin)(116.5,6){}
	\drawcurve[AHnb=0](113,2.5)(114,4)(115.5,4)(117,2.5)

  	\drawedge(7,8){}
  	\drawedge(8,9){}
  	\drawedge(8,10){}
\end{picture}
\caption{From parity games to simple stochastic games}\label{fig_red}
\end{center}
\end{figure}
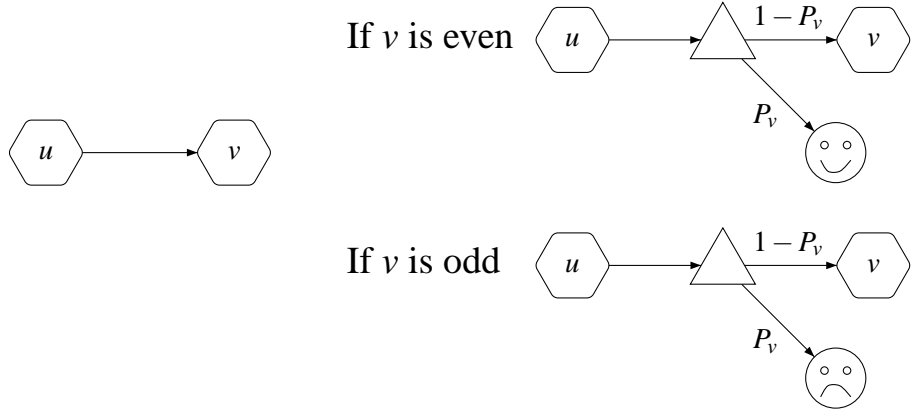

We present a pictorial description of the reduction in Figure~\ref{fig_red}:
for each edge $(u,v) \in E$, we consider the simple gadget, 
if $p(v)$ is even (resp. odd), that has an edge to the sink $\vw$ (resp. $\vl$) 
with probability $P_v$, and follows the original edge otherwise 
with probability $1 - P_v$.
Hexagonal vertices can be either Eve's or Adam's 
and triangle vertices are random vertices. 
$\vw$ will be depicted by a smiling face and $\vl$ by a sad face.

The new arena simulates the initial arena,
and additionally features two absorbing vertices $\vw$ and $\vl$.
To simulate a transition from $u$ to $v$, the new arena 
includes a random vertex that follows the transition with high probability $1 - P_v$
or stops the game by going to $\vw$ or $\vl$ with small probability $P_v$.
The intuition is that if $v$ has even priority, then Eve is rewarded to visit it
by having a small yet positive chance of winning,
and symmetrically if $v$ has odd priority for Adam.

Playing forever in $\Red(G)$, the outcome of the play will be favorable
for Eve (\textit{i.e} reach $\vw$) 
if she manages to see even priorities many times.
Furthermore, the reward a player receives for visiting a vertex with good priority
must depend on this priority:
seeing a very small even priority gives more chance to win than a higher one.
Indeed, if different priorities are seen infinitely often, 
the outcome of the play must be in favor of the parity of the lowest priority.
This leads to the following assumptions on the probabilities $P_v$'s.

\medskip\noindent{\bf Assumptions on the transition function.} 
We consider the following assumptions on $P_v$'s:
$$\sum_{v \in V} P_v \leq \frac{1}{6} \quad (A_0)$$
and for all $v \in V$, let $J_{\textrm{odd}}^{>v}  = \{ u \mid p(u) \textrm{ odd}, p(u) > p(v)\}$ and 
$J_{\textrm{even}}^{>v} = \{ u \mid p(u) \textrm{ even}, p(u) > p(v)\}$:
$$\sum_{u \in J_{\textrm{odd}}^{>v}} P_u \leq \frac{2}{3} \cdot P_v \qquad (A_1) \qquad \qquad
\sum_{u \in J_{\textrm{even}}^{>v}} P_u \leq \frac{2}{3} \cdot P_v \qquad (A_2)$$

We provide the reader with intuitions on the three assumptions $(A_0) - (A_2)$.
The assumption $(A_0)$ ensures that probabilities are small enough such that
plays in $\Red(G)$ will last enough to take into account the priorities seen
infinitely often, and not only the first ones.
The assumptions $(A_1)$ and $(A_2)$ ensure that if $v$ has the lowest priority 
and is seen infinitely often, no matters how often higher priorities are seen,
the outcome will only depend on the priority of $v$ and not on the others.

We present a sequence of properties to prove correctness of the reduction 
given the three assumptions $(A_0) - (A_2)$ hold.

First note that the set $\set{\vw,\vl}$ is reached with probability $1$,
since at each step there is a positive probability to reach it.
Another remark is that there is a one-to-one correspondence 
between strategies in $G$ and $\Red(G)$,
so we identify strategies in $G$ or $\Red(G)$.

We will prove that for all $v \in V$, 
Eve wins from $v$ if and only if $\valeve(\Reach(\vw))(v) \geq \frac{1}{2}$.

\begin{figure}
\begin{center}
\begin{picture}(100,20)(0,-10)
	\rpnode[arcradius=1](v)(20,0)(6,5){$v_0$}
	\rpnode[arcradius=1](i)(60,0)(6,5){$v_l$}
	\rpnode[arcradius=1](j)(75,10)(6,5){$v_{l+1}$}
  
	\drawedge[curvedepth=5,dash={2.5}0](v,i){}
	\drawqbpedge(i,90,j,-180){}
	\drawedge[curvedepth=15,dash={2.5}0](j,i){}
\end{picture}
\caption{General shape of a play in $2$-player game where Eve and Adam play positionally}
\label{fig_2play}
\end{center}
\end{figure}
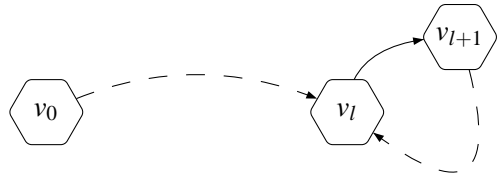

Thanks to Theorem~\ref{memdet}, there are memoryless optimal strategies 
in both games:
from now on, we consider $\strateve$ and $\stratadam$ two memoryless strategies.
The \textit{key} property is that the resulting play $\pi = \seq{v_0, v_1, \dots}$
has a simple shape (shown in Figure~\ref{fig_2play}):
the play consists in a simple path $\Pa$ from $v_0$ to $v_l$,
and then a simple cycle $\C$ is executed forever.
Let $c$ be the lowest priority infinitely visited,
$$\pi = \seq{v_0, v_1, \dots, v_{l-1}} \cdot \seq{v_l, v_{l+1}, \dots, v_{l+q-1}}^{\omega}$$
where $p(v_{l+1}) = c$, $\Pa = \set{v_0, v_1, \dots v_{l-1}}$ and
$\C = \set{v_l, v_{l+1}, \dots, v_{l+q-1}}$ are pairwise disjoint.

We now consider the corresponding situation in $\Red(G)$:
the random walk $\hpi = \play(v_0,\sigma,\tau)$ mimics $\pi$
until it takes an edge to $\vw$ or $\vl$, stopping the game.
We denote by $\hpi_i$ the random variable for the $i$\textsuperscript{th} vertex
of $\hpi$.
Since the starting vertex and the strategies are fixed, 
we will abbreviate $\Pro_{v_0}^{\strateve,\stratadam}$ by $\Pro$.

We consider the possible different scenarios. 
There are two possibilities to reach $\vw$ or $\vl$: 
the first is to reach it during the first $l$ steps, 
\textit{i.e} during the simple path, 
the second is to reach it after that, 
\textit{i.e} during the simple cycle,
after the simple path has been crossed. 

\medskip\noindent{\bf Notations for events.} 
We define, for $v \in \set{\vw,\vl}, k,j \geq 0$ the following measurable events.
\begin{itemize}
	\item  The event $\Reachgen{v}{j}$ denotes that $v$ has been reached within $j$ steps, \textit{i.e},
$\Reachgen{v}{j} = \set{\play \mid \hpi_j = v}$
(note that this is equivalent to $\set{\play \mid \exists i \leq j, \hpi_i = v}$).
	\item The event $\Crossgen{j}$ denotes that neither $\vw$ nor $\vl$ has been reached within $j$ steps, \textit{i.e}, \\
$\Crossgen{j} = \set{\play \mid \hpi_j \notin \set{\vw,\vl}}$.
	\item The event $\Rpath{v}$ denotes that $v$ has been reached within $l$ steps, \textit{i.e}, $\Rpath{v} = \Reachgen{v}{l}$.
	\item The event $\Cpath$ denotes that neither $\vw$ nor $\vl$ 
has been reached within $l$ steps, \textit{i.e}, $\Cpath = \Crossgen{l}$.
	\item The event $\Rloop{v}{k}$ denotes that $v$ has been reached 
within $l + k \cdot q$ steps, \textit{i.e}, \\
$\Rloop{v}{k} = \Reachgen{v}{l + k \cdot q}$:
intuitively, $v$ has been reached either during the path or 
one of the $k$ first crossings of the loop.
	\item The event $\Cloop{k}$ denotes that neither $\vw$ nor $\vl$ has been reached
within $l + k \cdot q$ steps,
\textit{i.e}, $\Cloop{k} = \Crossgen{l + k \cdot q}$:
intuitively, during the path and the $k$ first crossings of the loop
neither $\vw$ nor $\vl$ has been reached.
\end{itemize}

\noindent We define the following probabilities:
\begin{itemize}
	\item $\alpha = \Pro(\Cpath)$ is the probability to cross the path $\Pa$;
	\item $\beta = \Pro(\Rloop{\vw}{1} \mid \Cpath)$ is the probability 
to reach $\vw$ while following the simple cycle $\C$ for the first time, 
assuming the path $\Pa$ was crossed, and similarly
	\item $\gamma = \Pro(\Rloop{\vl}{1} \mid \Cpath)$.
\end{itemize}

\noindent We take two steps:
the first step is to approximate $\Pro(\Reach(\vw))$ and $\Pro(\Reach(\vl))$
using $\alpha$, $\beta$ and $\gamma$,
and the second is to make use of assumptions $(A_0) - (A_2)$
to evaluate $\alpha$, $\beta$ and $\gamma$.

\medskip\noindent{\bf Approximations for $\Pro(\Reach(\vw))$ and $\Pro(\Reach(\vl))$.}
We rely on the following four properties.

\begin{property}\label{prop1} For $k \geq 1$, we have 
$\Pro(\Rloop{\vw}{k} \mid \Cloop{k-1}) = \beta$
and similarly \\
$\Pro(\Rloop{\vl}{k} \mid \Cloop{k-1}) = \gamma$.
\end{property}
\begin{proof}
Since $\strateve$ and $\stratadam$ are memoryless, 
the random walk $\hpi$ is ``memoryless'': 
from $v_l$, crossing the loop for the first time or 
for the $k$-th time will give the same probability to escape to $\vw$ or $\vl$.
\hfill\qed
\end{proof}

\begin{property}\label{prop2} We have, for all $k \geq 1$, $\Pro(\Cloop{k-1} \mid \Cpath) = (1 - (\beta + \gamma))^{k-1}$.
\end{property}
\begin{proof}
By induction on $k \geq 1$. 
The case $k = 1$ follows from $\Cloop{0} = \Cpath$. 
Let $k > 1$:
$$\begin{array}{lll}
\multicolumn{3}{l}{\Pro(\Cloop{k} \mid \Cpath)} \\
\quad & = & \Pro(\Cloop{k} \mid \Cloop{k-1}) \cdot \Pro(\Cloop{k-1} \mid \Cpath) \\
\quad & = & (1 - \beta - \gamma) \cdot \Pro(\Cloop{k-1} \mid \Cpath)
\end{array}$$
The first equality is a restatement 
and the second is a result of Property~\ref{prop1}.
We conclude thanks to the induction hypothesis.
\hfill\qed
\end{proof}

\begin{property}\label{prop3} We have $\Pro(\Reach(\vw) \mid \Cpath) = \frac{\beta}{\beta + \gamma}$ and similarly
$\Pro(\Reach(\vl) \mid \Cpath) = \frac{\gamma}{\beta + \gamma}$.
\end{property}

A simple intuition on this calculation is by referring to a ``looping'' game.
Eve and Adam play a game divided in possibly infinitely many rounds.
Each round corresponds to cross the loop once: while doing so,
Eve wins with probability $\beta$, Adam wins with probability $\gamma$ and
the round is a draw otherwise, with probability $1 - (\beta + \gamma)$.
In case of a draw, the game goes on another round.
Once a player won, the game is stopped, which corresponds to reach $\vw$ or $\vl$.
In this game, Eve wins with probability $\frac{\beta}{\beta + \gamma}$
and Adam with probability $\frac{\gamma}{\beta + \gamma}$.

\begin{proof}
We have the following equalities:
$$\begin{array}{rcl}
\multicolumn{3}{l}{\Pro(\Reach(\vw) \mid \Cpath)} \\
\quad & = & \sum_{k = 1}^{\infty} \Pro(\Rloop{\vw}{k} \cap \Cloop{k-1} \mid \Cpath) \\[1ex]
\quad & = & \sum_{k = 1}^{\infty} \Pro(\Rloop{\vw}{k} \mid \Cloop{k-1}) \cdot \Pro(\Cloop{k-1} \mid \Cpath) \\[1ex]
\quad & = & \sum_{k = 1}^{\infty} \beta \cdot (1 - (\beta + \gamma))^{k-1} \\[1.5ex]
\quad & = & \frac{\beta}{\beta+\gamma}
\end{array}$$
The disjoint union 
$\Reach(\vw) \cap \Cpath = \uplus_{k \geq 1} (\Rloop{\vw}{k} \cap \Cloop{k-1})$
gives the first equality.
The second is a restatement, the third equality follows from Property~\ref{prop2}
and Property~\ref{prop1}.
The other equality is achieved by the same proof, replacing $\vw$ by $\vl$ and
using Property~\ref{prop1} accordingly.
\hfill\qed
\end{proof}

\begin{property}\label{prop4} We have 
$\Pro(\Reach(\vw)) \geq \alpha \cdot \frac{\beta}{\beta + \gamma}$ and similarly
$\Pro(\Reach(\vl)) \geq \alpha \cdot \frac{\gamma}{\beta + \gamma}$.
\end{property}

The intuition behind these two equalities is that we try to ignore what happens
while crossing the path, as reaching either $\vw$ or $\vl$
is not correlated to the priorities seen infinitely often.
In this context, the multiplicative constant $\alpha$ stands 
for the loss due to crossing the path.
As soon as the path is crossed, what happens next will be correlated to 
the priorities seen infinitely often along the play.
We will see that the value of the looping game described above captures 
the outcome of the parity game.

\begin{proof} We have the following equalities:
$$\begin{array}{rcl}
\Pro(\Reach(\vw)) & = & \Pro(\Rpath{\vw}) + \Pro(\Reach(\vw) \cap \Cpath) \\
& = & \Pro(\Rpath{\vw}) + \Pro(\Cpath) \cdot \Pro(\Reach(\vw) \mid \Cpath) \\
& \geq & \Pro(\Cpath) \cdot \Pro(\Reach(\vw) \mid \Cpath) \\
& = & \alpha \cdot \frac{\beta}{\beta + \gamma}
\end{array}$$
From the disjoint union
$\Reach(\vw) = \Rpath{\vw} \uplus (\Reach(\vw) \cap \Cpath)$
follows the first equality.
The second is restatement, the inequality is straightforward, 
the following equality is a restatement 
and the last equality follows from Property~\ref{prop3} and definition of $\alpha$.
The other claim is achieved by the same proof, replacing $\vw$ by $\vl$ and
using~\ref{prop3} accordingly.
\hfill\qed
\end{proof}

\medskip\noindent{\bf Approximations for $\alpha$, $\beta$ and $\gamma$.}
Note that for $i \geq 1$, we have
$\Pro(\hpi_i \in \set{\vw,\vl} \mid \hpi_{i-1} \notin \set{\vw,\vl}) = P_{v_i}$,
which follows from the construction of $\Red(G)$: taking an escape edge comes with probability $P_{v_i}$.

\begin{property}\label{prop5}
Given the assumption $(A_0)$ is satisfied, 
we have $\alpha \geq \frac{5}{6}$.
\end{property}

Intuitively, this property means that the loss due to crossing the path is 
bounded by a constant.

\begin{proof}
Since the path $\Pa$ is simple, each vertex is visited at most once.
Let $I = \set{i \mid v_i \in \Pa}$ 
be the set of vertices visited by this path. Then 
$$1 - \alpha = \sum_{i \in I} \Pro(\hpi_{i-1} \not\in \set{\vw,\vl} \cap \hpi_i \in \set{\vw,\vl}) \leq \sum_{i \in I} P_{v_i} \leq \frac{1}{6}$$
The first equality follows from the disjoint union:
$$\begin{array}{lll}
\multicolumn{3}{l}{\Play \setminus \Cpath} \\
\quad & = & \Rpath{\vl} \cup \Rpath{\vw} \\
\quad & = & \uplus_{i \in I} (\hpi_{i-1} \not\in \set{\vw,\vl} \cap \hpi_i \in \set{\vw,\vl})
\end{array}$$
The last inequality follows from assumption $(A_0)$.
\hfill\qed
\end{proof}

\begin{property}\label{prop6}
Given the assumptions $(A_1)$ and $(A_2)$ are satisfied, 
if Eve wins the play $\play(v_0,\sigma,\tau)$ in $G$, 
then we have the following inequalities:
$$\begin{array}{lccc}
(1) & \beta & \geq & P_{v_{l+1}} \\
(2) & \gamma & \leq & \frac{2}{3} \cdot P_{v_{l+1}}
\end{array}$$
and similarly 
if Adam wins the play $\play(v_0,\sigma,\tau)$ in $G$, 
then we have the following inequalities:
$$\begin{array}{lccc}
(1) & \gamma & \geq & P_{v_{l+1}} \\
(2) & \beta & \leq & \frac{2}{3} \cdot P_{v_{l+1}}.
\end{array}$$
\end{property}

Intuitively, this property means that if Eve wins in the parity game,
then the looping game is winning for her 
with probability more than $\frac{2}{3}$,
and similarly if Adam wins in the parity game,
then the looping game is winning for him 
with probability more than $\frac{2}{3}$.

\begin{proof}
We prove inequalities in both cases simultaneously.
\begin{enumerate}
	\item It relies on the fact that the loop starts by getting to $v_{l+1}$,
\textit{i.e} that either, if Eve wins:
$\hpi_{l+1} = \vw \cap \Cpath \subseteq \Rloop{\vw}{1} \cap \Cpath$,
or if Adam wins:
$\hpi_{l+1} = \vl \cap \Cpath \subseteq \Rloop{\vl}{1} \cap \Cpath$.
	\item 
Assume Eve wins, let $J = \set{i \mid v_i \in \C \wedge p(v_i) \mbox{ odd}}$ be
the set of vertices with odd priority visited by the loop. Then
$$\gamma = \sum_{i \in J} \Pro(\hpi_{i-1} \neq \vl \cap \hpi_i = \vl) 
\leq \sum_{i \in J} P_{v_i}$$
$(A_1)$ allows to conclude, 
since $v_{l+1}$ has the lowest priority of the loop, thus
$J \subseteq J_{\textrm{odd}}^{>v_{l+1}}$.
Similarly, if Adam wins, the same proof using assumption $(A_2)$ concludes.
\end{enumerate}
\hfill\qed
\end{proof}

It follows from Property~\ref{prop4},~\ref{prop5} and ~\ref{prop6} that 
under assumptions $(A_0) - (A_2)$,
we have the desired equivalence: Eve wins in $G$ if and only if
$\Pro(\Reach(\vw)) \geq \frac{1}{2}$ in $\Red(G)$.

\begin{theorem} 
Under the three assumptions $(A_0) - (A_2)$, we have:
for all $2$-player arenas $G$ equipped with parity objective $\Parity(p)$,
$\Red(G)$ equipped with reachability condition $\Reach(\vw)$ 
is a simple stochastic game with the stopping property, 
and for all $v \in V$, Eve wins for the parity condition from $v$ in $G$ 
if and only if $\valeve(\Reach(\vw))(v) \geq \frac{1}{2}$ in $\Red(G)$.
\end{theorem}

\medskip\noindent{\bf Transition probabilities.} 
We now present transition probabilities satisfying the assumptions $(A_0)-(A_2)$
that can be described with $O(\log(n))$ bits.
Let $p:V \to \N$ the priority function in $G$, 
we first build an equivalent parity function $p'$. 
We sort vertices with respect to $p$ and define the following \textit{monotone} mapping: 
\begin{itemize}
	\item the lowest priority becomes either $4$ if it is even or $5$ if odd;
	\item proceeding from the lowest to the greatest, a vertex is assigned 
the lowest integer greater than the last integer used, matching its parity.
\end{itemize} 
This ensures that all priorities are distinct, 
and the highest priority is at most $2n+2$.
Then, apply the reduction $\Red$ with $P_j = \frac{1}{2^j}$.
We argue that the probability transition function satisfies $(A_0)-(A_2)$.
We have
\[
\sum_{v \in V} P_v \leq \frac{1}{2^4} + \frac{1}{2^5} + \ldots
= \frac{1}{8}  \leq \frac{1}{6}
\]
Hence $(A_0)$ is satisfied.
For all $v \in V$,
\[
\sum_{u \in J_{\textrm{odd}}^{>v}} P_u \leq 
P_v \cdot \bigg(\frac{1}{2} + \frac{1}{2 \cdot 2^2} + \frac{1}{2 \cdot 2^4} + \ldots\bigg) = 
P_v \cdot \frac{1}{2} \cdot \frac{4}{3} = \frac{2}{3}\cdot P_v
\]
Hence $(A_1)$ is satisfied and a similar argument holds for $(A_2)$.
Hence we have the following result.

\begin{theorem}
$\Red$ is a polynomial-time reduction from $2$-player parity games
to simple stochastic games.
Furthermore, for all $2$-player parity games $(G,\Parity(p))$, 
the size of the stochastic arena $\Red(G)$ is $O(|E| \cdot \log(|V|))$.
\end{theorem}

\section{Reducing parity games to simple stochastic games}

In this section, we discuss related works.
Deciding the winner in parity games is equivalent to the model-checking problem
of modal mu-calculus.
A reduction from model-checking games to simple stochastic games 
was defined in~\cite{Stir99}.
Another reduction, using a discounted mu-calculus, from concurrent parity games 
to concurrent discounted games was presented in~\cite{AHM03}.
Our intend was to propose a direct and simple reduction from $2$-player parity games
to simple stochastic games.
In the first subsection, we discuss its efficiency 
compared to the previously known three step reduction. 
In the second subsection, we use remarks from~\cite{Con93-DMTCS} 
to prove that solving stochastic parity, mean-payoff, discounted-payoff games 
as well as simple stochastic games is in $\UP \cap \co\UP$.

\subsection{Discounting the discounted}

In this subsection we present the classical sequence of reductions:
from $2$-player parity games to $2$-player mean-payoff games~\cite{Jur98},
from $2$-player mean-payoff games to $2$-player discounted-payoff games~\cite{ZP96},
and from $2$-player discounted games to simple stochastic games~\cite{ZP96}.

\smallskip\noindent{\bf Parity games to mean-payoff games.}
A $2$-player parity game with $n$ vertices and $d$ different priorities can 
be reduced in polynomial time to a $2$-player mean payoff game on the same arena
using rewards from the set $\{-n^d,\dots,n^d\}$, such that Eve wins the parity 
game if and only if the value of Eve in the mean-payoff game is at least $0$~\cite{Jur98}.

\smallskip\noindent{\bf Mean-payoff games to discounted-payoff games.}
A $2$-player mean payoff game with $n$ vertices whose reward function ranges from $-B$ to $B$ can be reduced in polynomial time to a discounted-payoff game on the same arena with discount factor $\lambda$ such that $\lambda \geq 1 - \frac{B}{4n^3}$ such that the value of Eve in the mean-payoff game 
is at least $0$ if and only if the value of Eve in the discounted-payoff game is at least $0$~\cite{ZP96}.

\smallskip\noindent{\bf Discounted-payoff games to simple stochastic games.}
A $2$-player discounted-payoff game with $n$ vertices can be reduced in polynomial time 
to a simple stochastic game using $n + m$ vertices including $m$ random vertices and 
$4\cdot m$ edges such that the value of Eve in the discounted-payoff 
game is at least $0$
if and only if the value of Eve 
in the simple stochastic game is at least $\frac{1}{2}$~\cite{ZP96}.
%
%
%
%
%
%

\smallskip\noindent{\bf Size of the resulting games.}
We now analyze the size of the games produced by the three step reduction.
Let $G$ a $2$-player parity game having $n$ vertices, $m$ edges and $d$ distinct priorities. 
The first reduction to a $2$-player mean payoff game yields a game with $n$ vertices, 
$m$ edges and rewards can be specified with $O(d \cdot \log(n))$ bits. 
The second reduction to a discounted-payoff game  yields a $2$-player 
game with $n$ vertices, $m$ edges, rewards specified with $O(d \cdot \log(n))$ bits 
and the discount  factor specified with $O(d \cdot \log(n))$ bits. 
Finally, the last reduction to a simple stochastic game yields a game with 
$n + m$ vertices, with $m$ random vertices, $4\cdot m$ edges and each probability of transition
specified with $O(d \cdot \log(n))$ bits, thus the size of the transition 
function is $O(m \cdot (\log(n+m) + d \cdot \log(n))) = O(m \cdot d \cdot \log(n))$.
Since $d$ is $O(n)$, in the worst case the size of the game obtained by 
the three step reduction is $O(m \cdot n \cdot \log (n))$.

\subsection{The complexity of stochastic games}

Another motivation to present a clean and direct reduction 
from $2$-player parity games to simple stochastic games
was to extend it from stochastic parity games to simple stochastic games.
As for the deterministic case, such a reduction is known, 
but again through stochastic mean-payoff and stochastic discounted-payoff,
and is more involved~\cite{CH08}.
Although we did not manage to adapt our proofs to extend our direct reduction
from stochastic parity games to simple stochastic games, we believe it is possible.
Our main difficulty is that the shape of a play, 
even if both players play positionally,
is no more a ``lasso-play''.
Indeed, even if the parity condition is satisfied with probability more than half, 
we cannot guarantee that an even priority will be visited 
within a linear number of steps.

In the remaining of this subsection, we gather several results and make 
two very simple observations of the result of Condon~\cite{Con93-DMTCS} to
prove that the decision problem for simple stochastic games is in $\UP \cap \co\UP$,
in a similar fashion to the proof of~\cite{Jur98}, which was 
stated for the simpler case of $2$-player discounted games.

\begin{figure}
\begin{center}
\begin{picture}(125,50)(0,0)
	\gasset{Nadjust=w}

	\node(parity)(15,35){parity}
	\node(mp)(47,45){mean-payoff}
	\node(dp)(97,45){discounted-payoff}

	\node(ssg)(120,18){simple stochastic games}

	\node(sparity)(0,16){stochastic parity}
	\node(smp)(30,0){stochastic mean-payoff}
	\node(sdp)(90,0){stochastic discounted-payoff}
	
  	\drawedge(parity,mp){[6]}
  	\drawedge(mp,dp){[9]}
  	\drawedge(dp,ssg){[9]}
  	\drawedge[linewidth=.3](parity,ssg){$\Red$}
  	\drawedge(sparity,smp){}
  	\put(17,8){[3]}
  	\drawedge(smp,sdp){[2]}
  	\drawedge(sdp,ssg){}
  	\put(110,8){[2]}

	\drawline[dash={1.5}0,AHnb=0](-15,27)(135,27)
	\put(0,28.5){$2$-player}
	\put(-1,23){stochastic}
\end{picture}
\caption{Reductions}
\label{red}
\end{center}
\end{figure}
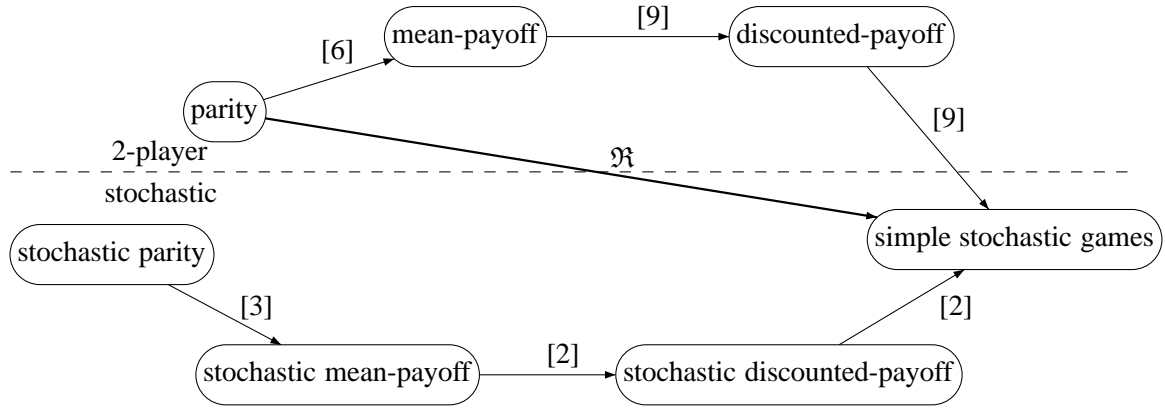

The reduction of stochastic parity to stochastic mean-payoff games 
was established in~\cite{CH08} and reduction of stochastic mean-payoff and
discounted games to simple stochastic games was established in~\cite{AM09}.
The Figure~\ref{red} summarizes all the reductions.
We now argue that simple stochastic games can be decided 
in $\UP \cap \co\UP$.

\smallskip\noindent{\bf Simple stochastic games in $\UP \cap \co\UP$.} 
First, it was shown in~\cite{Con92-IC} that simple stochastic games with 
arbitrary rational transition probabilities can be reduced in polynomial time to 
stopping simple stochastic games where random vertices have 
two outgoing edges each with probability half.
Second, it follows from the result of~\cite{Con93-DMTCS} that the value vector of a \textit{stopping} simple stochastic game 
is the unique solution of the following equations set:
$$\left\{
\begin{array}{ll}
\valeve(\vw) = 1 \\
\valeve(\vl) = 0 \\
\valeve(v) = \max_{(v,v') \in E} \mathrm{val_E}(v') & \text{ if } v \in \VE \\
\valeve(v) = \min_{(v,v') \in E} \mathrm{val_E}(v') & \text{ if } v \in \VA \\
\valeve(v) = \sum_{(v,v') \in E} \delta(v,v') \cdot \mathrm{val_E}(v') & \text{ if } v \in \VR 
\end{array}
\right.$$
Hence an algorithm can guess the value vector and check that 
it is actually the solution of the equation set. 
To prove the desired result we need to show that the guess is of polynomial size and 
the verification can be achieved in polynomial time.
It follows from~\cite{Con92-IC} that for simple stochastic games with $n$ vertices
and all probabilities one half, the values are of the form $p / q$, 
where $p,q$ are integers, $0 \leq p,q \leq 4^{n-1}$. 
Hence the length of the guess is at most $n \cdot \log (4^{n-1}) = O(n^2)$, 
which is polynomial.
Thus the guess is of polynomial size and the verification 
can be done in polynomial time.
The unique solution implies that simple stochastic games are in $\UP$, and 
the $\co\UP$ argument is symmetric. 
Along with the reductions of~\cite{AM09,CH08} we obtain the following 
result.

\begin{theorem}[Complexity of stochastic games]
For all stochastic arenas, for all objectives $\phi$ such that $\phi$ is a parity, mean-payoff, discounted-payoff or 
reachability objective, the decision problem of whether $\valeve(\phi)(v) \geq q$, for a rational 
number $q$  is in $\UP \cap \co\UP$.
\end{theorem}

\smallskip\noindent{\bf Acknowledgements.} 
The second author thanks Florian Horn for his guidance and support
during the preparation of this paper.

\bibliographystyle{eptcs}
\bibliography{people,short,papers}

\begin{thebibliography}{1}
\providecommand{\bibitemdeclare}[2]{}
\providecommand{\urlprefix}{Available at }
\providecommand{\url}[1]{\texttt{#1}}
\providecommand{\href}[2]{\texttt{#2}}
\providecommand{\urlalt}[2]{\href{#1}{#2}}
\providecommand{\doi}[1]{doi:\urlalt{http://dx.doi.org/#1}{#1}}
\providecommand{\bibinfo}[2]{#2}

\bibitemdeclare{inproceedings}{AHM03}
\bibitem{AHM03}
\bibinfo{author}{Luca {d}e Alfaro}, \bibinfo{author}{Thomas~A. Henzinger} \&
  \bibinfo{author}{Rupak Majumdar} (\bibinfo{year}{2003}):
  \emph{\bibinfo{title}{Discounting the Future in Systems Theory}}.
\newblock In: {\sl \bibinfo{booktitle}{International Colloquium on Automata,
  Languages and Programming, ICALP}}, pp. \bibinfo{pages}{1022--1037},
  \doi{10.1007/3-540-45061-0\_79}.

\bibitemdeclare{inproceedings}{AM09}
\bibitem{AM09}
\bibinfo{author}{Daniel Andersson} \& \bibinfo{author}{Peter~Bro Miltersen}
  (\bibinfo{year}{2009}): \emph{\bibinfo{title}{The Complexity of Solving
  Stochastic Games on Graphs}}.
\newblock In: {\sl \bibinfo{booktitle}{International Symposium on Algorithms
  and Computation, ISAAC}}, pp. \bibinfo{pages}{112--121},
  \doi{10.1007/978-3-642-10631-6\_13}.

\bibitemdeclare{article}{CH08}
\bibitem{CH08}
\bibinfo{author}{Krishnendu Chatterjee} \& \bibinfo{author}{Thomas~A.
  Henzinger} (\bibinfo{year}{2008}): \emph{\bibinfo{title}{Reduction of
  stochastic parity to stochastic mean-payoff games}}.
\newblock {\sl \bibinfo{journal}{Information Processing Letters, IPL}}
  \bibinfo{volume}{106}(\bibinfo{number}{1}), pp. \bibinfo{pages}{1--7},
  \doi{10.1016/j.ipl.2007.08.035}.

\bibitemdeclare{article}{Con92-IC}
\bibitem{Con92-IC}
\bibinfo{author}{Anne Condon} (\bibinfo{year}{1992}):
  \emph{\bibinfo{title}{{T}he {C}omplexity of {S}tochastic {G}ames}}.
\newblock {\sl \bibinfo{journal}{Information and Computation}}
  \bibinfo{volume}{96}(\bibinfo{number}{2}), pp. \bibinfo{pages}{203--224},
  \doi{10.1016/0890-5401(92)90048-K}.

\bibitemdeclare{inproceedings}{Con93-DMTCS}
\bibitem{Con93-DMTCS}
\bibinfo{author}{Anne Condon} (\bibinfo{year}{1993}):
  \emph{\bibinfo{title}{{O}n {A}lgorithms for {S}imple {S}tochastic {G}ames}}.
\newblock In: {\sl \bibinfo{booktitle}{{A}dvances in {C}omputational
  {C}omplexity {T}heory}}, {\sl \bibinfo{series}{DIMACS Series in Discrete
  Mathematics and Theoretical Computer Science}}~\bibinfo{volume}{13},
  \bibinfo{publisher}{American Mathematical Society}, pp.
  \bibinfo{pages}{51--73}, \doi{10.1.1.46.6099}.

\bibitemdeclare{article}{Jur98}
\bibitem{Jur98}
\bibinfo{author}{Marcin Jurdzi{\'n}ski} (\bibinfo{year}{1998}):
  \emph{\bibinfo{title}{Deciding the Winner in Parity Games is in {UP} $\cap$
  co-{UP}}}.
\newblock {\sl \bibinfo{journal}{Information Processing Letters, IPL}}
  \bibinfo{volume}{68}(\bibinfo{number}{3}), pp. \bibinfo{pages}{119--124},
  \doi{10.1016/S0020-0190(98)00150-1}.

\bibitemdeclare{article}{Stir99}
\bibitem{Stir99}
\bibinfo{author}{Colin Stirling} (\bibinfo{year}{1999}):
  \emph{\bibinfo{title}{Bisimulation, Modal Logic and Model Checking Games}}.
\newblock {\sl \bibinfo{journal}{Logic Journal of the IGPL}}
  \bibinfo{volume}{7}(\bibinfo{number}{1}), pp. \bibinfo{pages}{103--124},
  \doi{10.1093/jigpal/7.1.103}.

\bibitemdeclare{incollection}{Tho97}
\bibitem{Tho97}
\bibinfo{author}{Wolfgang Thomas} (\bibinfo{year}{1997}):
  \emph{\bibinfo{title}{{Languages, Automata, and Logic}}}.
\newblock In \bibinfo{editor}{G.~Rozenberg} \& \bibinfo{editor}{A.~Salomaa},
  editors: {\sl \bibinfo{booktitle}{Handbook of Formal Languages}},
  chapter~\bibinfo{chapter}{7}, \bibinfo{volume}{3, Beyond Words},
  \bibinfo{publisher}{Springer}, pp. \bibinfo{pages}{389--455},
  \doi{10.1.1.38.8643}.

\bibitemdeclare{article}{ZP96}
\bibitem{ZP96}
\bibinfo{author}{Uri Zwick} \& \bibinfo{author}{Mike Paterson}
  (\bibinfo{year}{1996}): \emph{\bibinfo{title}{The complexity of mean payoff
  games on graphs}}.
\newblock {\sl \bibinfo{journal}{Theoretical Computer Science, TCS}}
  \bibinfo{volume}{158}, pp. \bibinfo{pages}{343--359},
  \doi{10.1016/0304-3975(95)00188-3}.

\end{thebibliography}

\end{document}